\documentclass[letterpaper,twocolumn,10pt]{article}
\usepackage{usenix2019_v3}

\usepackage{tikz}
\usepackage{amsmath}

\usepackage{subcaption}

\begin{document}
	%-------------------------------------------------------------------------------
	
	%don't want date printed
	\date{}
	
	\title{\Large \bf VOICE-ZEUS: Impersonating Zoom's E2EE-Protected Static Media and Textual Communications via Simple Voice Manipulations}

	%for single author (just remove % characters)
	\author{
		{\rm Mashari Alatawi}\\
		Texas A\&M University
		\and
		{\rm Nitesh Saxena}\\
		Texas A\&M University
		% copy the following lines to add more authors
		% \and
		% {\rm Name}\\
		%Name Institution
	} % end author

	\maketitle
	
	%-------------------------------------------------------------------------------
	\begin{abstract}
		The authentication ceremony plays a crucial role in verifying the identities of users before exchanging messages in end-to-end encryption (E2EE) applications, thus preventing impersonation and man-in-the-middle (MitM) attacks. Once authenticated, the subsequent communications in E2EE apps benefit from the protection provided by the authentication ceremony. However, the current implementation of the authentication ceremony in the Zoom application introduces a potential vulnerability that can make it highly susceptible to impersonation attacks. The existence of this vulnerability may undermine the integrity of E2EE, posing a potential security risk when E2EE becomes a mandatory feature in the Zoom application. In this paper, we examine and evaluate this vulnerability in two attack scenarios, one where the attacker is a malicious participant and another where the attacker is a malicious Zoom server with control over Zoom's server infrastructure and cloud providers. Our study aims to comprehensively examine the Zoom authentication ceremony, with a specific focus on the potential for impersonation attacks in static media and textual communications. We simulate a new session injection attack on Zoom E2EE meetings to evaluate the system's susceptibility to simple voice manipulations. Our simulation experiments show that Zoom's authentication ceremony is vulnerable to a simple voice manipulation, called a VOICE-ZEUS attack, by malicious participants and the malicious Zoom server. In this VOICE-ZEUS attack, an attacker creates a fingerprint in a victim's voice by reordering previously recorded digits spoken by the victim. We show how an attacker can record and reorder snippets of digits to generate a new security code that compromises a future Zoom meeting. Therefore, the mandatory use of E2EE in Zoom can create a vulnerability, enabling a malicious participant or malicious Zoom server to impersonate hosts and perform zoombombing attacks with offensive content. We conclude that stronger security measures are necessary during the group authentication ceremony in Zoom to prevent impersonation attacks.
	\end{abstract}

	\section{Introduction}
	End-to-end encryption (E2EE) stands as a fundamental security feature, ensuring that digital communication between users is thoroughly encrypted and exclusively accessible to the intended recipients. The essence of E2EE lies in its ability to ensure that only the intended recipients possess the decryption keys, rendering the content inaccessible to any third party, including service providers. The primary purpose of E2EE is to thwart any unauthorized access or interception of communication, thereby fortifying the privacy and confidentiality of the users involved. Moreover, E2EE's encryption mechanism provides a robust defense against eavesdropping or espionage activities perpetrated by governments or other malicious actors.
	
	In all E2EE applications, the authentication ceremony stands as a pivotal step in establishing secure communication among users or devices. Its primary goal is to validate the identities of the parties involved and facilitate the secure exchange of encryption keys. Ensuring the integrity and security of E2EE communication requires that only the intended recipients have access to decrypted messages. Consequently, the authentication ceremony plays a crucial role in securing all subsequent communications. Following the verification of identities and the exchange of keys, the authentication ceremony results in the creation of a secure channel. Through this fortified pathway, data, messages, static media, and textual communications can be transmitted freely and safely, alleviating concerns about interception by malicious actors. The significance of the authentication ceremony extends across diverse domains, be it in personal interactions, financial transactions, or corporate communications. It establishes itself as the foundational bedrock of trust within the digital realm, forming a perimeter of confidence that assures users that their shared information is protected and accessible only to those possessing the correct encryption keys. 
	
	Video conferencing platforms, such as Zoom, have revolutionized communication in the digital era, providing an intuitive solution for virtual meetings and collaborative activities \cite{zoom}. The COVID-19 pandemic accelerated the reliance on Zoom, transforming how people communicate and collaborate in business, education, and personal settings due to restrictions on physical meetings and gatherings \cite{pelosi2021zoomimpact}. However, the increased use of video conference systems has led to a surge in security concerns. There have been reports of the Chinese government engaging in espionage and security activities on U.S. soil, with alleged instances of Chinese security officials spying on Zoom calls and harassing Chinese dissidents \cite{china2023zoomspy}. These cases have drawn attention to the vulnerabilities and risks associated with video conferences, as they can become targets of governments or malicious actors seeking unauthorized access, eavesdropping, or disruption of communication. In this context, ensuring the confidentiality and integrity of communication has become paramount, with security measures such as E2EE becoming a critical tool in protecting against unauthorized access and espionage activities by governments or malicious actors.
	
\begin{figure}[t]
		\centering
		\includegraphics[width=.35\textwidth]{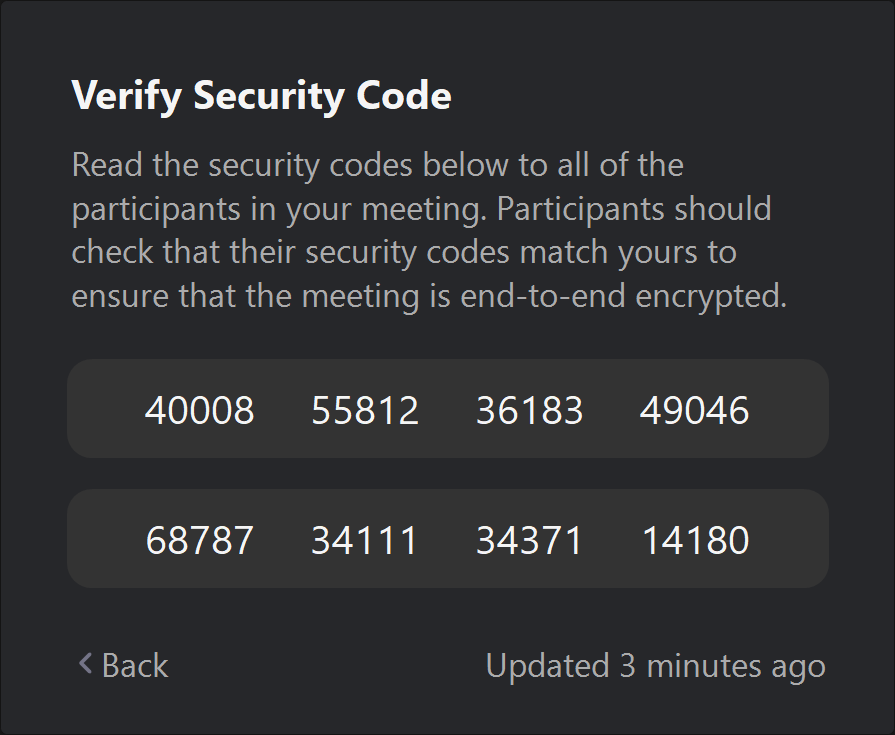}
		\caption{Authentication ceremony in Zoom} \label{fig1}
		%\vspace{-3mm}
	\end{figure}

	Despite its merits, E2EE in Zoom faces certain challenges. One of these challenges is the potential for attacks on the underlying authentication ceremony of Zoom. In E2EE applications, users typically need to verify the authenticity of the individual they are communicating with. This verification is commonly accomplished through methods like cross-checking a contact's fingerprint or scanning a QR code displayed on the recipient's device. Such methods empower users to authenticate the genuineness of their communication partner, establishing confidence that they are indeed engaging with the intended individual and not an impersonator. In practice, the Zoom application incorporates an authentication ceremony that includes a security code consisting of numerical digits, which users use to verify the authenticity of the host (see Figure \ref{fig1}). However, the implementation of a secure authentication ceremony is not without its challenges, as it must be resilient against various threats, including man-in-the-middle (MitM) attacks and other forms of impersonation. Ensuring the authenticity of keys and the identity of communication partners is essential.
	
	One notable aspect of concern is the absence of an out-of-band (OOB) channel, such as Short Messaging Service (SMS) or email, for the secure exchange of the security code during the authentication ceremony in Zoom. Consequently, the only way for the host to verify the code with participants is by reading it aloud for the verification to occur in-band \cite{zoome2ee2020}. This practice, while essential for the authentication process, does introduce a potential vulnerability, as it may expose the authentication ceremony to exploitation by malicious actors. Adversaries could exploit this vulnerability by illicitly capturing the host's voice without consent during the authentication ceremony. They may employ external devices or third-party recording software to clandestinely record the host's vocalized security code. This captured data could then be employed to impersonate the previous host in subsequent Zoom E2EE meetings, potentially compromising the integrity and security of the communication environment.
	
	In a future Zoom E2EE meeting session, a malicious adversary could effortlessly engage in simple voice manipulations by assembling previously recorded audio snippets. This could be used to generate a new security code, mimicking the voice of the previous host. Consequently, the adversary can deceive any participant into accepting the session, leading them to believe that it genuinely originates from the legitimate host and successfully complete the authentication ceremony. Upon the successful completion of the authentication ceremony, all forms of communication within the session, including text-based conversations and screen-sharing presentations, are considered authenticated. The adversary can abuse the chat and screen-sharing functionalities, displaying obscene content on behalf of the victim host. Such actions have the potential to significantly damage the victim host's reputation, and in severe cases, may expose them to legal repercussions for the inappropriate content shared during the session.
	
	To demonstrate the potential risks of impersonation attacks on Zoom E2EE meetings, we simulated simple voice manipulations on the current implementation of the authentication ceremony in Zoom. This manipulation, termed a VOICE-ZEUS attack, involves rearranging previously recorded spoken numbers by an attacker, resulting in the creation of a vocal fingerprint that matches that of the victim. The authentication ceremony in Zoom currently uses only numeric representation for its security code and does not use any OOB channel to exchange the security code (see Figure \ref{fig1}). In our study, we created two separate Zoom E2EE meetings: one generated by the legitimate host and the other by a malicious actor attempting to impersonate the previous host. Text-to-speech (TTS) samples and Free Spoken Digit Dataset (FSDD) \cite{fsdd} were used to read the security code aloud during the authentication ceremony of the first Zoom E2EE meeting. Our experiments involved multiple attempts to perform the attack, and the host's voice was recorded using various methods, including the Audacity software \cite{audacity} and a smartphone device. The outcomes of our experiments expose the vulnerability in the current implementation of the authentication ceremony in Zoom, emphasizing the imperative for robust security measures to protect against impersonation attacks in static media and textual communications.
	
	\noindent \textbf{Our Contributions:} Our contributions are as follows:
	
	\begin{itemize}
		\item We highlight a vulnerability in the current implementation of the Zoom authentication ceremony, demonstrating its susceptibility to a VOICE-ZEUS attack. This flaw enables an attacker to inject a new Zoom E2EE session into legitimate sessions, allowing them to impersonate a previous host.
		\item Through simulation, we illustrate a new session injection attack targeting group-based E2EE protocols like Zoom. In this scenario, a legitimate Zoom E2EE session is generated by a victim acting as the meeting host, while an attacker, posing as a legitimate participant, records the host's voice for future use. Subsequently, the attacker creates an injected Zoom E2EE session, reordering the previously recorded audio snippets to impersonate the victim's voice during the authentication ceremony and introduce a new security code. Our research underscores the risk of adversaries participating innocuously as legitimate participants, exploiting third-party software or external devices to record a host's voice. This recorded voice can later be used for impersonation in subsequent Zoom E2EE meetings, particularly in static media and textual communications.
		\item We provide a thorough evaluation of our attack, both in terms of its feasibility and its effectiveness. We conducted an objective evaluation to quantitatively measure the similarity of a reordered voice to the original voice. The results conclusively establish the susceptibility of the current Zoom authentication ceremony to impersonation attacks, where attackers can effortlessly record a host's voice in one session and employ a VOICE-ZEUS attack in a future session. These findings underscore the imperative of implementing robust security measures to mitigate the risk of such attacks in Zoom E2EE meetings.
	\end{itemize}

	\noindent \textbf{Attack Demonstration:} A video demonstration of the attack is available at https://sites.google.com/view/zoom-attack/home

	\section{Background}

	\subsection{Zoom Video Communications}
	Zoom Video Communications, commonly known as Zoom \cite{zoom}, has gained massive popularity since its inception in 2011, thanks to its user-friendly interface, simplicity, versatility, and ease of use. The COVID-19 pandemic has immensely accelerated the adoption of Zoom as a preferred communication tool among businesses, educational institutions, and individuals worldwide, due to the need for social distancing and remote work \cite{pelosi2021zoomimpact}. This has highlighted the necessity of a viable communication solution, and Zoom has emerged as an efficient and trustworthy option that fulfills this requirement. Its intuitive interface and rich set of features make it a top choice for virtual communication among diverse users \cite{hanselzoom2use,jiangzoom2use,pelosi2023zoomfeatures}. One of its pivotal features is its virtual meeting capability, enabling participants from diverse locations to seamlessly connect and cooperate, making it an indispensable tool for remote work, telecommuting, global partnerships, and more. Moreover, Zoom's recording feature enables users to record meetings or webinars for future reference or sharing with absentees, while screen sharing enables participants to share their screens with others, facilitating more collaborative and interactive meetings.
	
	However, it is worth mentioning that Zoom has faced some challenges and controversies as well \cite{wan2020zoom,villasenor2020zoom}. At the outset of the pandemic, apprehensions surfaced regarding the effectiveness of Zoom's security and privacy measures, as reported cases of "zoombombing" involved disruptive unauthorized participants or insiders with legitimate access introducing inappropriate content \cite{zoombombing2021}. These incidents have highlighted the inherent risks of relying on digital platforms such as Zoom to conduct academic activities. Furthermore, this trend has exacerbated concerns surrounding potential threats to academic freedom and privacy, specifically concerning the phenomenon known as the "zoomification" of academia, and the associated issues of data privacy, surveillance, and censorship \cite{zoomification2021}. With increased use, video conferencing platforms such as Zoom have become appealing targets for eavesdropping and espionage activities by governments and other malicious actors. These incidents have brought into question the platform's overall security posture and prompted Zoom to address the identified vulnerabilities and fortify its security features. Consequently, Zoom has implemented robust security measures, such as E2EE, to protect the privacy and integrity of meetings and data exchanged during Zoom sessions and to reinforce the overall security posture of the platform \cite{zoome2ee2020}.

	\subsection{End-to-End Encryption}
	The implementation of E2EE has become crucial in modern communication systems to protect against unauthorized access and data breaches. It has emerged as an effective security protocol for ensuring the confidentiality and security of sensitive information shared during virtual meetings. E2EE guarantees that only the intended recipients of the communication can decrypt and access the data shared during the communication, making it particularly attractive to companies and organizations handling sensitive data. This prevents unauthorized access, interception, or interpretation of the data by malicious actors. This robust method encrypts the communication at the sender's device and decrypts it at the receiver's device, effectively preventing unauthorized entities, including service providers and attackers, from intercepting or tampering with the data during transit. E2EE's ability to provide this level of security makes it an essential tool for protecting sensitive information in virtual meetings. 
	
	Recently, Zoom incorporated E2EE in its meetings, leveraging public key cryptography where the host generates encryption keys and distributes them to the meeting participants \cite{zoome2ee2020}. The keys are then used to encrypt and decrypt the data shared during the meetings, making it inaccessible to outside parties, including Zoom itself. To use this feature, users must enable E2EE in their account settings before setting up the meeting. However, E2EE meetings in Zoom are not compatible with certain features like cloud recording and live transcription \cite{zoomwhitepaper}, as they require access to unencrypted data, thus posing data security risks to users. Cloud recordings, for example, could be vulnerable to attack by malicious servers (such as the service provider Zoom), which could access the information stored in them. Thus, while E2EE offers many advantages in data security, online communication platforms need to ensure compatibility with all features and assure users that their sensitive information is secure at all times.
	
	E2EE applications currently operate in an opportunistic mode, wherein the security assurances they provide depend on users trusting the service provider. This mode establishes secure connections between parties without mutual authentication \cite{langley2009opportunistic,herzberg2016can}. Despite its opportunistic nature, these applications offer significant privacy advantages over conventional communication methods like email and standard instant messaging \cite{unger2015sok,ermoshina2016end}. They effectively protect users from passive attackers, such as government surveillance agencies, seeking to eavesdrop on connections and collect sensitive information.
	
	While opportunistic E2EE mode protects against passive MitM attacks, it leaves the system susceptible to active MitM attacks. These attacks may be perpetrated by rogue service providers or by compromising the server itself \cite{herzberg2016can,yadav2022automatic}. The critical security challenge arises from the dependence on provider-supplied keys in these E2EE applications. During the key exchange service, a malicious or hacked server might simply launch an attack known as a key substitution attack, which compromises the entire E2EE system. This key substitution attack involves replacing a legitimate key with a fake one controlled by the malicious server. Consequently, without proper authentication of the encryption keys, users cannot ensure that they are indeed communicating with the intended recipients and not with an impersonator or malicious attacker.
	
	Similar to other E2EE applications, Zoom employs an opportunistic E2EE mode that enables the creation of a secure communication channel between two parties without authenticating the other party. Consequently, the E2EE security of the Zoom application heavily relies on the authentication ceremony, a process that verifies the security code of encryption keys shared by participants. This ceremony allows Zoom users to verify that they are communicating exclusively with the intended parties during Zoom communications, thereby preventing impersonation and MitM attacks. Therefore, users must compare and verify the security code through this ceremony to establish complete end-to-end security. By adopting this approach, users can detect and thwart potential attack attempts, ensuring the secure sharing of sensitive information during virtual meetings.
	
	\subsection{Authentication Ceremony}
	
	\subsubsection{An Overview of Authentication Ceremony}
	In E2EE applications, an authentication ceremony is a process that allows users to verify each other's identities before exchanging sensitive information. This ceremony is essential in ensuring that only authorized individuals are granted access to confidential data, and it also helps prevent impersonation and MitM attacks. Typically, the authentication ceremony involves a series of steps that all users must follow. These steps may include verifying each other's public keys or verifying a conversation's fingerprint. The exact process may vary depending on the application, but the goal is always the same: to establish trust and ensure that all users are who they claim to be.
	
	The authentication ceremony holds a crucial role in guaranteeing the authenticity of the communication process in E2EE applications. This task verifies the identity of involved parties and ensures that users possess the correct encryption keys for intended recipients, thwarting potential threats from fake keys provided by malicious entities. Its primary function is to act as a safeguard against impersonation and MitM attacks, thereby upholding the integrity of the entire communication. At present, the authentication ceremony in E2EE applications relies heavily on users actively participating in manually verifying and confirming the identities of their communication partners \cite{alatawi2023sok}. This involvement includes actions such as in-person QR code scanning or verbal confirmation of key fingerprints during a phone call. However, the reliance on manual user intervention introduces inherent weaknesses that attackers can exploit to launch impersonation and MitM attacks.
	
	A notable vulnerability arises from the possibility of human errors, with users overlooking or improperly verifying the authenticity of key fingerprints. Factors like time pressure, fatigue, or distractions further compound these errors, escalating the risk of successful impersonation and MitM attacks. Moreover, the existing manual authentication ceremony faces usability and robustness issues, potentially rendering E2EE communication insecure \cite{herzberg2016can,schroder2016signal,shirvanian2015security,shirvanian2017pitfalls,vaziripour2017you}. Usability problems, such as the absence of OOB channels for exchanging key fingerprints during the authentication ceremony, elevate the risk of unintentionally establishing connections with malicious actors posing as legitimate communication partners. Through successful impersonation, an attacker gains the ability to intercept and manipulate communication without the knowledge of the legitimate users, compromising the confidentiality and integrity of exchanged data.
	
	Therefore, the authentication ceremony is indispensable in dispelling doubts regarding key verification. It mandates all communication parties to verify each other's key fingerprint before exchanging encrypted messages. The key fingerprint, serving as a unique identifier for the specific encryption key in use, ensures that both parties are genuinely communicating with each other and not falling victim to an attacker. The user's understanding and active participation are pivotal for the successful completion of the authentication ceremony. Based on the match or mismatch of encryption keys, users must decide whether to proceed with the communication or discontinue it. Consequently, good design becomes paramount, as the effectiveness of the authentication ceremony relies on users taking specific steps and being fully aware of the consequences of neglecting them.
	
	\subsubsection{The Authentication Ceremony in Zoom}
	To ensure that virtual meetings held on Zoom are private, users need to compare the security code of the meeting's host to confirm that they have received the correct public key for the meeting's host. In Zoom E2EE meetings, as described in the Zoom whitepaper \cite{zoomwhitepaper}, participants are asked to validate the authenticity of the leader's (host) public key (\textit{IVK}$_l$) for each E2EE meeting, thereby thwarting any potential impersonation or MitM attacks. The security code for a Zoom meeting host is computed using the following equation:

	{
		\small{
			\begin{equation}
				\begin{split}
					&Digits(SHA256(SHA256(\\
					&''Zoombase-1-ClientOnly-MAC-SecurityCode'')\\
					&\|SHA256(\textit{IVK}_{l})))
				\end{split}
		\end{equation}}
	}
	
	%\noindent $Digits(SHA256(SHA256(''Zoombase-1-ClientOnly-MAC-SecurityCode'')\|SHA256(\textit{IVK}_{l})))$

\begin{figure*}[t]
		\centering
		\includegraphics[width=.80\linewidth]{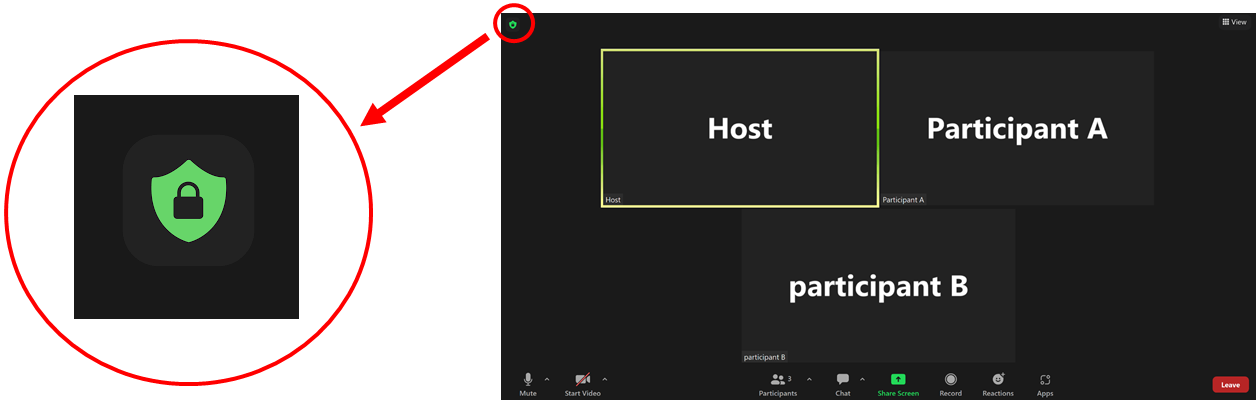}
		\caption{Using the green shield icon to find and locate the authentication ceremony in Zoom.} \label{fig2}
		%\vspace{-7mm}
	\end{figure*}

	Each application provides a different user interface to facilitate the authentication ceremony, and Zoom uses a user-friendly interface to locate its authentication ceremony. After joining the meeting, participants can check for the green shield icon in the upper-left corner of the meeting window to access the authentication ceremony (see Figure \ref{fig2}). The Zoom client interface presents the meeting host's security code in a more comprehensible format than the hexadecimal hash, as it is designed to be human-readable. As shown in Figure \ref{fig1}, the security code is composed of a string of 40 decimal digits, which are split into 8 blocks of 5-digit numbers. To ensure authenticity and prevent impersonation and MitM attacks, each participant's Zoom client in the E2EE meeting independently computes the security code using the (\textit{IVK}$_l$) that is initially employed in the handshake protocol. Regrettably, the Zoom application does not utilize an OOB channel to exchange the security code during the authentication ceremony. Consequently, the only recourse available to the meeting host may involve verbally communicating the security code, with all participants listening and verifying that their clients display the same security code. Nonetheless, the use of only numeric codes without an OOB channel renders the system vulnerable to VOICE-ZEUS attacks. In such attacks, an adversary can record the user's voice uttering the digits 0 through 9 during a prior Zoom meeting and subsequently reorder the snippets to generate a specific code that compromises a future Zoom meeting.
	
	After successfully completing the authentication ceremony, the attacker can continue impersonating the previous host and play inappropriate or offensive content such as obscene video or abusive language during the meeting, which is known as a zoombombing attack. Such incidents can be very disruptive and traumatizing, especially if they involve hate speech or graphic images. In addition to causing emotional harm, impersonation attacks can also compromise the privacy and security of the meeting, as the attacker may have access to sensitive information or recordings. The zoombombing phenomenon became a widespread issue during the early days of the COVID-19 pandemic when many people started using Zoom for virtual meetings and classes \cite{zoombombing2021}. However, with the current implementation of the authentication ceremony in Zoom, the risk of such attacks remains as long as Zoom continues to use numeric code only for its security code and does not use an OOB channel for the exchange of authentication codes.

	\subsection{Related Work}
	In recent years, E2EE has become increasingly popular as a method of ensuring the confidentiality and privacy of online communications. Zoom, one of the most widely used video conferencing platforms, recently introduced E2EE for its users \cite{zoome2ee2020}. However, the security of this feature has been called into question by several researchers who have identified a number of potential vulnerabilities and attacks. In a study by Isobe and Ito \cite{isobe2021security}, the authors conducted a thorough security evaluation of the E2EE feature in Zoom, examining various attacks. They discussed an attack where colluding participants share the meeting key with a malicious Zoom server, compromising meeting confidentiality. The authors explored additional attacks by exploiting vulnerabilities in the E2EE protocol. They proposed impersonation attacks by malicious participants in group meetings, impersonation by the malicious Zoom server using binding information from previous meetings, collusion of participants with the malicious Zoom server to fully impersonate legitimate users, and collusion-enabled impersonation attacks when multiple users share a device. They also highlighted security concerns such as the use of vulnerable encryption mode and potential denial of service by the malicious Zoom server. 
	
	Our work shares similarities with the above work \cite{isobe2021security} in terms of focusing on assessing the security of Zoom's E2EE feature. However, our study differs from their work as we specifically focus on examining the implementation of the authentication ceremony in the Zoom application. We highlight a weakness in the authentication ceremony and provide evidence of a particular type of attack that can exploit this vulnerability. We specifically explore the vulnerability of the authentication ceremony arising from the absence of using OOB channels, relying solely on voice to announce the security code. The authors in \cite{isobe2021security} acknowledge that their findings are based on the Zoom whitepaper \cite{zoomwhitepaper} and they have analyzed only the cryptographic protocol of E2EE. They did not implement or test their proposed attacks. In contrast, our work proposes a specific attack on the authentication ceremony in Zoom's E2EE feature. We have also implemented a demo attack to demonstrate the feasibility of our proposed attack.
	
	In a study by Kagan et al. \cite{kagan2020zooming}, privacy risks in virtual meetings were highlighted, revealing the ease of extracting personal information like face images, age, gender, usernames, and even full names from publicly available images. This extracted data poses security and privacy threats online and in the real world, especially when cross-referenced with social network data. Similarly, our work focuses on the privacy and security risks in virtual meetings. We demonstrate an attack where malicious users can collect sensitive personal data, such as voice samples, by recording the victim's voice during the authentication ceremony. Through simple voice manipulations, the attacker can create a new security code in the host's voice, allowing them to impersonate the meeting host in future meetings and carry out malicious activities.
	
	Similar to our work, the authors in \cite{shirvanian2014wiretapping} use voice reordering attacks to challenge the assumption that the voice channel, used for validating Short Authenticated Strings (SAS), provides integrity and source authentication in voice over the Internet (VoIP) communications. They report on automated SAS voice imitation MitM attacks that compromise the security of Crypto Phones. While their focus is on MitM attacks and short codes like SAS, our work employs simple voice manipulations with a longer numeric code in an authentication ceremony. We show a real impersonation attack rather than a real-time MitM attack, making our approach more practical.

	\section{Our Impersonation Attack}
	\label{main_attack}
	In this section, we will demonstrate an attack on group-based E2EE protocols (such as a group meeting in Zoom) under the threat model as described in Subsection \ref{threadmodel}. We explore the security implications of relying solely on the audible transmission of numbers during the authentication ceremony in group communication protocols, including Zoom. We demonstrate that this approach can pose a significant security vulnerability, as it is susceptible to new session injection attacks that leverage simple voice manipulations.

	\subsection{Overview of the Attack}
	\label{ourattack}
	The impersonation attack is a serious security threat in group-based E2EE protocols, such as Zoom. The attack involves injecting a new session into legitimate communication sessions between two or more parties, with the aim of creating a new session with different parameters or goals. From a security perspective, E2EE is a protocol, and the authentication ceremony is a crucial component of it. Therefore, our aim is to demonstrate that E2EE can be compromised by breaking its authentication ceremony. Here are the steps of the attack that must be executed successfully to compromise the group E2EE in Zoom: 
	
	\begin{enumerate}
		\item Recording the host's voice as they speak only digits, aiming to collect audio snippets of all ten digits (from 0 to 9) for future use:
		\begin{enumerate}
			\item The attacker can join a legitimate Zoom E2EE meeting as a legitimate participant and record the host's voice during the authentication ceremony when the host verbally announces the numerical security code.
			\item The attacker could be a malicious Zoom server with access to Zoom recordings stored in the cloud from default Zoom meetings (non-E2EE meetings) that were previously recorded by the host. Subsequently, the attacker extracts only the audio snippets containing the digits.
			\item The malicious Zoom server could potentially infiltrate default Zoom meetings that do not employ E2EE, simply by using meeting IDs known to the Zoom server. Subsequently, the attacker can gain access to parts of the host's voice during conversations where numbers are spoken and record them.
		\end{enumerate}
		\item After capturing the necessary voice snippets, the malicious participant or the malicious Zoom server can proceed to create a new Zoom E2EE meeting session and falsely claim it was initiated by the previous host. The attacker may employ social engineering or phishing attacks to deceive and invite other participants to join the new session. 
		\item Generating the new security code in the previous host's voice is accomplished by performing a VOICE-ZEUS attack, where snippets of legitimate audio signals from previous recordings are cut and pasted.
		\item The attacker can ask participants to compare and verify the new security code for the new meeting during the authentication ceremony.
		\item Once the attacker successfully completes the authentication ceremony, all forms of communication within the session, including text-based interactions and screen-sharing presentations, are considered authenticated.
		\item The attacker can then proceed with the session, utilizing chat and screen-sharing features available within the meeting to disseminate offensive language, racial slurs, or even threaten specific individuals. This can result in harm to participants, damage the reputation of the victim host, and create a hostile environment.
	\end{enumerate}
	
\begin{figure*}[t]
		\centering
		\includegraphics[width=.80\linewidth]{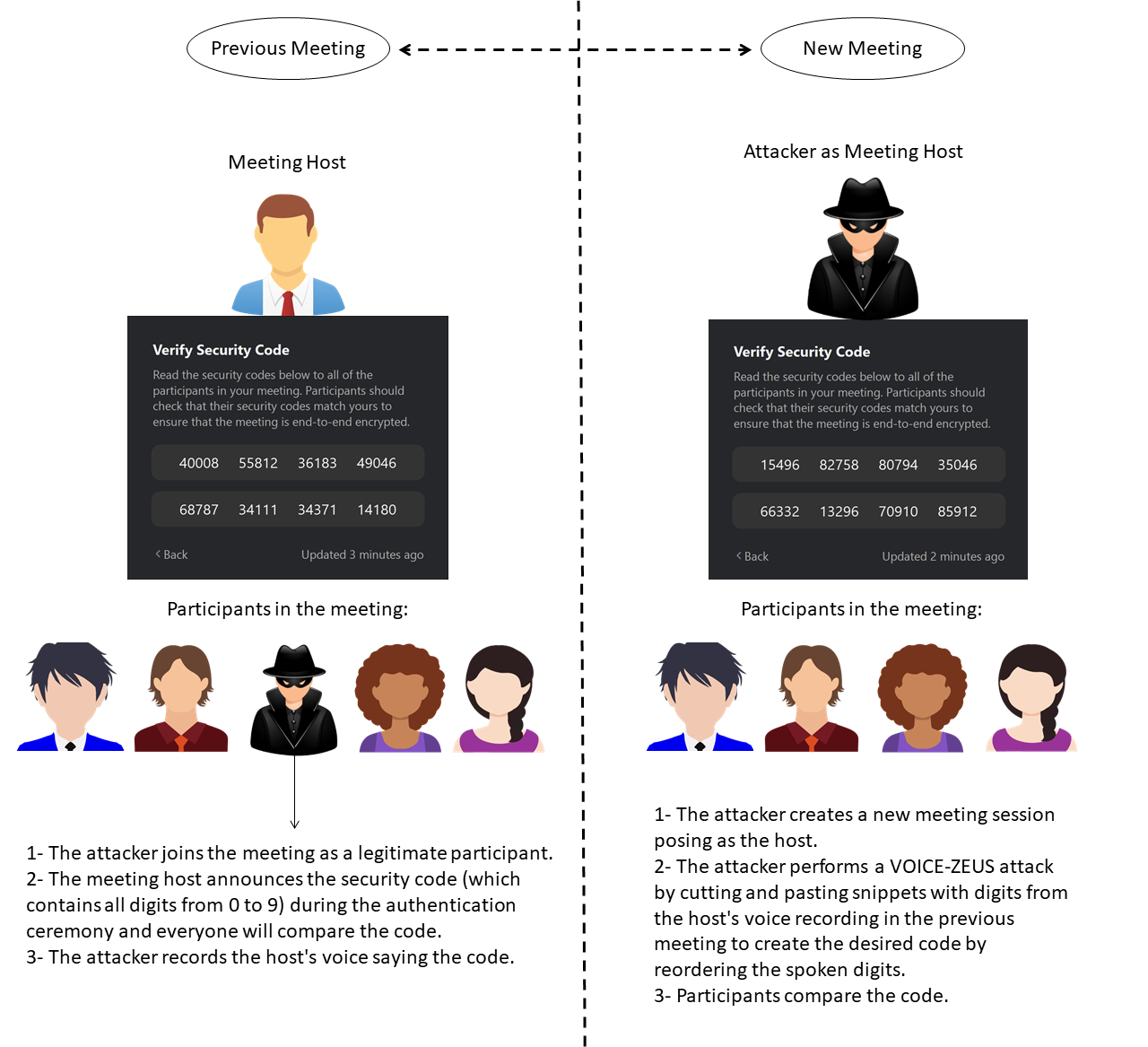}
		\caption{A high-level overview of our attack} \label{fig3}
		%\vspace{-5mm}
	\end{figure*}

	To demonstrate the effectiveness of this attack, we will perform an attack on a Zoom E2EE meeting and use our proposed VOICE-ZEUS attack to generate a security code in the voice of the host. A high-level overview of our attack is provided in Figure \ref{fig3}.

	\subsection{Threat Model and Attacker Assumptions}
	\label{threadmodel}
		
	Our study examines the capabilities of an attacker at two levels:
	
	\subsubsection{Malicious Meeting Participant}
	We consider a scenario where the attacker is a malicious participant in the Zoom E2EE meeting, possessing the ability to join it as a legitimate participant. In this situation, the attacker is not required to make any substantial effort to gain access to and exploit the meeting. A malicious participant can easily implement such an attack by joining a Zoom E2EE meeting and recording the host's voice as they speak the security code aloud during the authentication ceremony. The attacker only needs to gain one-time access to a legitimate E2EE meeting during the authentication ceremony stage. The security code spoken during the authentication ceremony comprises all digits from 0 to 9, providing the attacker with all the audio snippets for these digits. The threat model is highly feasible, especially with the growing adoption of video conferencing technology in diverse sectors such as education, business, and healthcare.
	
	\subsubsection{Malicious Zoom Server}
	We make the assumption that the attacker has control over Zoom's server infrastructure and its cloud providers, enabling the attacker to act as either a rogue server or a compromised server, also known as a malicious Zoom server. The malicious Zoom server is capable of joining any default Zoom meeting, which is a non-E2EE meeting, as a participant, using the meeting ID known to the Zoom infrastructure. This meeting ID is a necessary precondition for joining a meeting, as described in the Zoom whitepaper \cite{zoomwhitepaper}. Additionally, the malicious Zoom server can leverage its ability to enforce server features, such as removing or adding a participant to a meeting or altering the display names of any meeting participant, to add itself to a meeting. The objective of the malicious Zoom server is to obtain access to the default Zoom meeting and record the host's voice during a conversation when saying digits ranging from 0 to 9. To accomplish this goal, the malicious Zoom server may need to join one or several sessions to ensure recording all the necessary digits. Because the default Zoom meeting lacks E2EE and an authentication ceremony, it is crucial to note that the attacker may not obtain all the digits, as there is no security code. However, some default Zoom meetings are automatically recorded on the cloud by hosts, thus providing the malicious Zoom server with access to all snippets for all digits from 0 to 9. The mandatory use of E2EE in Zoom in the future could introduce a potential vulnerability, enabling a malicious Zoom server to impersonate any Zoom meeting host and gain access to sensitive information.
	
	\subsection{Attacking Scenarios}
	In a malicious Zoom meeting scenario, we assume that the attacker as a host has disabled the video feature, leaving only the audio voice functional for the duration of the meeting. The role of the host in managing an audio-only meeting is pivotal, especially in situations where the attacker claims issues with the video feature. When initiating an audio-only meeting on Zoom, the attacker possesses the capability to customize the meeting settings before it begins. This includes the option to disable video, ensuring that participants only communicate using audio channels. This functionality is easily accessible through the host's Zoom account settings, allowing for a seamless transition to audio-only mode during the meeting.
	
	Once the video is disabled and the authentication ceremony is completed, it becomes easy for the attacker to limit communication to non-vocal means. The attacker may claim issues with their audio or video and proceed with the meeting, relying solely on the chat function for textual communications. Additionally, the attacker can utilize the screen-sharing feature as a valuable tool for visual communication, displaying images, slides, or other visual aids instead of verbal content.
	
	In all the following scenarios, we assume that the host (the attacker) will follow the same steps outlined in Subsection \ref{ourattack} to execute the attack. Additionally, the attacker is assumed to have the ability to disable the audio/video feature and conduct meetings using only the chat and screen-sharing features.
	
	\subsubsection{Educational Setting}
	In a virtual parent-teacher meeting on Zoom, an attacker acquires meeting details and impersonates a teacher. A new Zoom session is set up with E2EE, and the teacher's voice is recorded from an earlier session or public event. Parents, hearing the familiar voice during the authentication ceremony, join the meeting under the false assumption that it is real. With access to the authenticated session, the impersonator manipulates the text conversation and delivers false or misleading information regarding the child's academic progress. They may provide inaccurate assessments, misrepresent the child's abilities, or even request sensitive personal information under the guise of updating records. These actions can lead to confusion, mistrust, and potential harm to the child's educational development or compromised privacy.
	
	In another scenario, a member of the student council with malicious intent infiltrated into an earlier Zoom E2EE meeting when the president discussed private school matters. Extracting voice snippets, the attacker orchestrates a deceptive E2EE session, crafting invitations mirroring the president's style. Assuming the president's identity, the attacker sends invites to council members for an unscheduled session on urgent matters. Trusting the seemingly credible host, members unknowingly join, becoming susceptible to a VOICE-ZEUS attack during verification. With control established, the attacker disseminates misinformation, causing chaos and eroding the president's credibility. Exploiting chat features, the attacker engages in disruptive zoombombing, playing inappropriate content and sending offensive messages while posing as the president.
	
	\subsubsection{Healthcare Setting}
	In a virtual appointment scenario, a patient schedules a Zoom E2EE meeting with their doctor to discuss medical conditions. However, an attacker intercepts the email and creates a fake meeting, impersonating the doctor. The patient joins the meeting unaware of the impersonation, and the attacker cuts and pastes snippets of the doctor's voice to complete the authentication ceremony. Once authenticated, the attacker can provide harmful medical advice, steal personal information, or engage in fraudulent activities. Text-based conversations are used to deceive the patient into sharing sensitive data, which can be exploited for various malicious purposes. This puts the patient's privacy and security at risk, potentially resulting in financial losses for both the patient and healthcare provider.
	
	In a healthcare scenario, a hospital organizes a virtual Zoom conference to discuss medical advancements. However, a malicious individual gains unauthorized access, impersonating a renowned medical expert. The impersonator records snippets of the expert's voice and sets up a deceptive E2EE session, matching the security code with the expert's voice to mislead participants. Healthcare professionals, thinking they're in an authoritative discussion, unknowingly accept the manipulated security code. The impersonator, having authenticated the session, selectively uses chat and screen-sharing to present fabricated research findings or advocate questionable medical practices. This deceitful conduct can compromise patient care, hinder medical progress, and potentially exploit attendees by soliciting personal information.
	
	\subsubsection{Other Settings}
	In a banking scenario, a customer schedules a Zoom E2EE meeting with their bank representative for loan consolidation. However, an attacker gains unauthorized access to the customer's email, impersonates the bank representative, and creates a fraudulent meeting. The attacker records voice snippets from a prior Zoom meeting, rearranging them to match the representative's voice for a new security code announcement. The customer joins the meeting, unaware of the impersonation, and the attacker guides them through an authentication ceremony where the reordered security code is presented as legitimate. With access to the authenticated session, the attacker engages in deceptive practices using textual conversation. They may claim to offer special loan terms or request sensitive financial information such as social security numbers, bank account details, or credit card information. The attacker can then use this information for identity theft, financial fraud, or unauthorized access to the customer's accounts.
	
	In a virtual sales presentation, a potential client schedules a Zoom E2EE meeting with a sales representative. However, a malicious party intercepts the email and creates a fake meeting, impersonating the sales representative. The client joins the meeting, unaware of the impersonation. The attacker has previously recorded snippets of the representative's voice from a different Zoom meeting or accessed previous recordings stored on the cloud. Employing these recorded voice snippets, the attacker creates a new Zoom E2EE meeting session and inserts the new security code to match the voice of the sales representative. During the authentication ceremony, the attacker asks the client to compare and verify the security code, successfully tricking the client into thinking the meeting is legitimate. Once authenticated, all forms of communication within the session, including text-based interactions and screen sharing, are considered authenticated. Taking advantage of the authenticated session, the attacker starts the sales presentation, but with malicious intentions. The attacker may provide false information, misleading proposals, or attempt to gather sensitive business data. These actions can lead to financial losses for the client and damage the reputation of the sales representative and their organization.

	\subsection{Attack Experiment Setup}
	To evaluate the feasibility of our attack, we conducted two Zoom E2EE meetings, each involving three users, with one acting as the host and two as participants. Our experimental setup consisted of a laptop computer, an iPhone device, and an Android device, with one device serving as the host and the other devices as participants. In the initial meeting session, we established a legitimate Zoom E2EE meeting using the iPhone device as the host. The laptop computer and Android device were used as participants, with the laptop computer serving as the attacker whose objective was to record the host's voice uttering numbers from 0 to 9 for future use. During the authentication ceremony, we generated an audio sample of the security code using a free online TTS generator \cite{ttsmp3} for experimental purposes. We used samples from three male English speakers and three female English speakers, representing US, British, and Australian accents. Here the attacker needs to capture the host's voice while uttering numbers, utilizing any available software that offers an audio recording service. It is assumed that the host has not granted permission for any participants to record the meeting. Consequently, the attacker will need to use third-party software or an external device, such as a smartphone, to record the host's voice. In our study, we utilized both methods to record the host's voice for all speaker samples. First, we utilized the Audacity software \cite{audacity} to record the host's voice during the authentication ceremony and export it as a WAV-format audio file. Second, we used an external device (Samsung Galaxy S21) to record the host's voice during the authentication ceremony, employing the Samsung Voice Recorder app \cite{voicerecorder} installed on the device to capture and export the voice as a WAV-format audio file.
	
	During the initial meeting session, the attacker's objective is to detect the presence of any digit in either the authentication ceremony or the regular conversation uttered by the host. It is obvious that all digits can be spoken during the security code presentation in the authentication ceremony. Therefore, the attacker only needs to activate the Audacity software before the host initiates the authentication ceremony to ensure recording all digits. For a different type of attacker, such as the malicious Zoom server, the attacker may need to record the entire Zoom meeting or acquire access to the Zoom recordings stored in the cloud. Subsequently, the attacker can extract only the snippets containing the digits from the host's voice for future use. In the second meeting session, we established a Zoom E2EE meeting as an attacking session, utilizing the laptop computer as the host of the meeting, with the primary objective of impersonating the previous host. We utilized the other devices (the iPhone device and the Android device) to participate in the meeting as legitimate participants. During the authentication ceremony, we executed the VOICE-ZEUS attack to impersonate a previous legitimate host by reordering all digits of the security code in his voice. In our experimental setup, we attempted the attack twelve times and captured the host's voice on six occasions using the Audacity software and six instances using the Samsung Galaxy S21 phone device. The process of conducting the attack involved the creation of two Zoom E2EE meetings on each attempt. We utilized samples of six speakers (TTS samples) to read out the security code during the authentication ceremony. The TTS samples were produced using a freely available online TTS generator \cite{ttsmp3}.
	
	To further investigate the feasibility of our attack, we expanded our experimental scope by incorporating speech samples from the FSDD dataset in addition to the TTS samples. The FSDD dataset contains audio recordings of single digits (0-9) spoken by six human speakers, saved in WAV files, and is available as an open-source voice dataset at \cite{fsdd}. In our expanded experiment, we followed the same set of steps as in our previous experimental setup. We recorded spoken digits in one Zoom meeting and reordered them in another Zoom meeting. During the initial meeting session, we introduced speech samples from the FSDD dataset to read out the security code during the authentication ceremony. The attacker's objective remained the same: capturing the host's voice while they uttered the numbers. To record the host's voice, we utilized both the Audacity software and an external device (Samsung Galaxy S21), enabling us to export the recorded audio as WAV files. In the subsequent meeting session, created as the attacking session, we performed a VOICE-ZEUS attack. This entailed rearranging all digits of the security code in the voice of a previous legitimate host during the authentication ceremony, achieved by cutting and pasting snippets of digits extracted from previously recorded audio files. Similar to the previous scenarios involving TTS-generated samples, we conducted twelve attempts of this extended experiment using human speech samples from the FSDD dataset. We were able to capture the host's voice in six instances using the Audacity software and another six instances using the Samsung Galaxy S21.

	\section{Evaluation}
	In line with the attack described in Section \ref{main_attack}, we simulated an attempt to impersonate a host within an injected Zoom session. To evaluate the effectiveness of this attack, we conducted a series of tests wherein we created two distinct Zoom E2EE meetings, captured the voice of the host in the first meeting, and subsequently executed a VOICE-ZEUS attack to impersonate a previous host by recreating the new security code in his voice.
	
\begin{figure*}[t]
		\begin{subfigure}{.30\textwidth}
			\centering
			\includegraphics[width=1\linewidth]{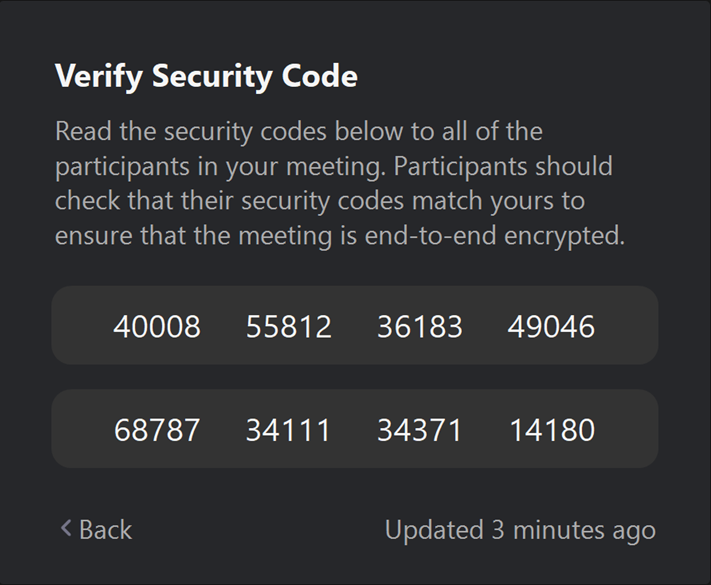}
			\caption{The previous host's security code}
			\label{fig4:host_security_code}
			%\vspace{-1mm}
		\end{subfigure}\hspace{35mm}%
		\begin{subfigure}{.45\textwidth}
			\centering
			\includegraphics[width=1\linewidth]{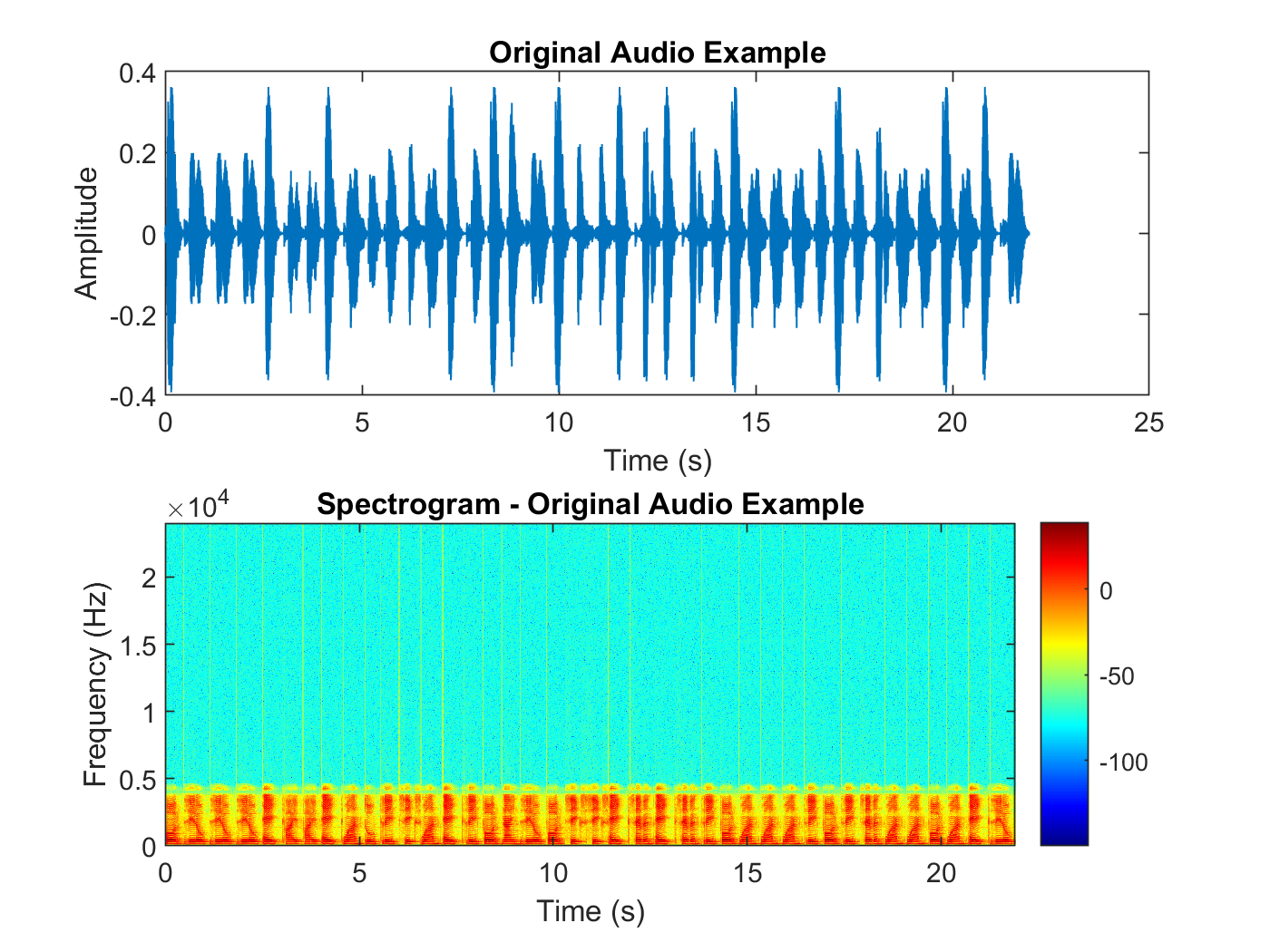}
			\caption{Plot the original audio wave in time and frequency domain}
			\label{fig4:plot_time_freq}
			%\vspace{-3mm}
		\end{subfigure}\hspace{35mm}%
		\begin{subfigure}{.30\textwidth}
			\centering
			\includegraphics[width=1\linewidth]{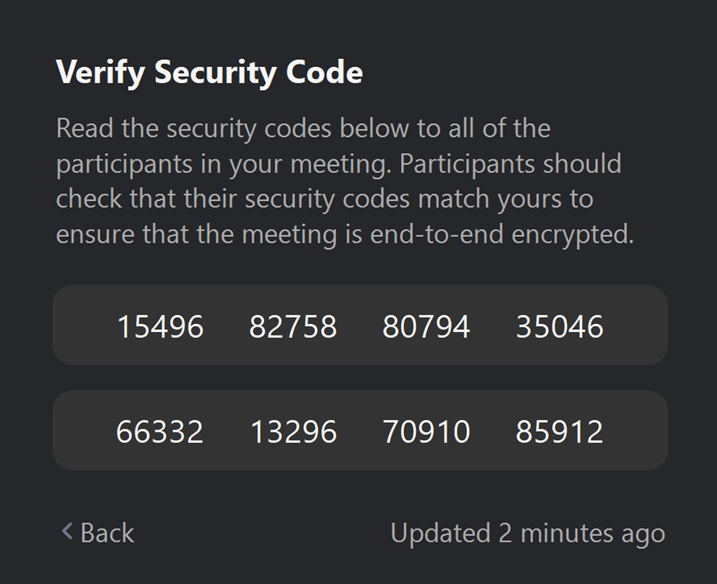}
			\caption{The current host's security code}
			\label{fig5:host_security_code}
			%\vspace{-1mm}
		\end{subfigure}\hspace{35mm}%
		\begin{subfigure}{.45\textwidth}
			\centering
			\includegraphics[width=1\linewidth]{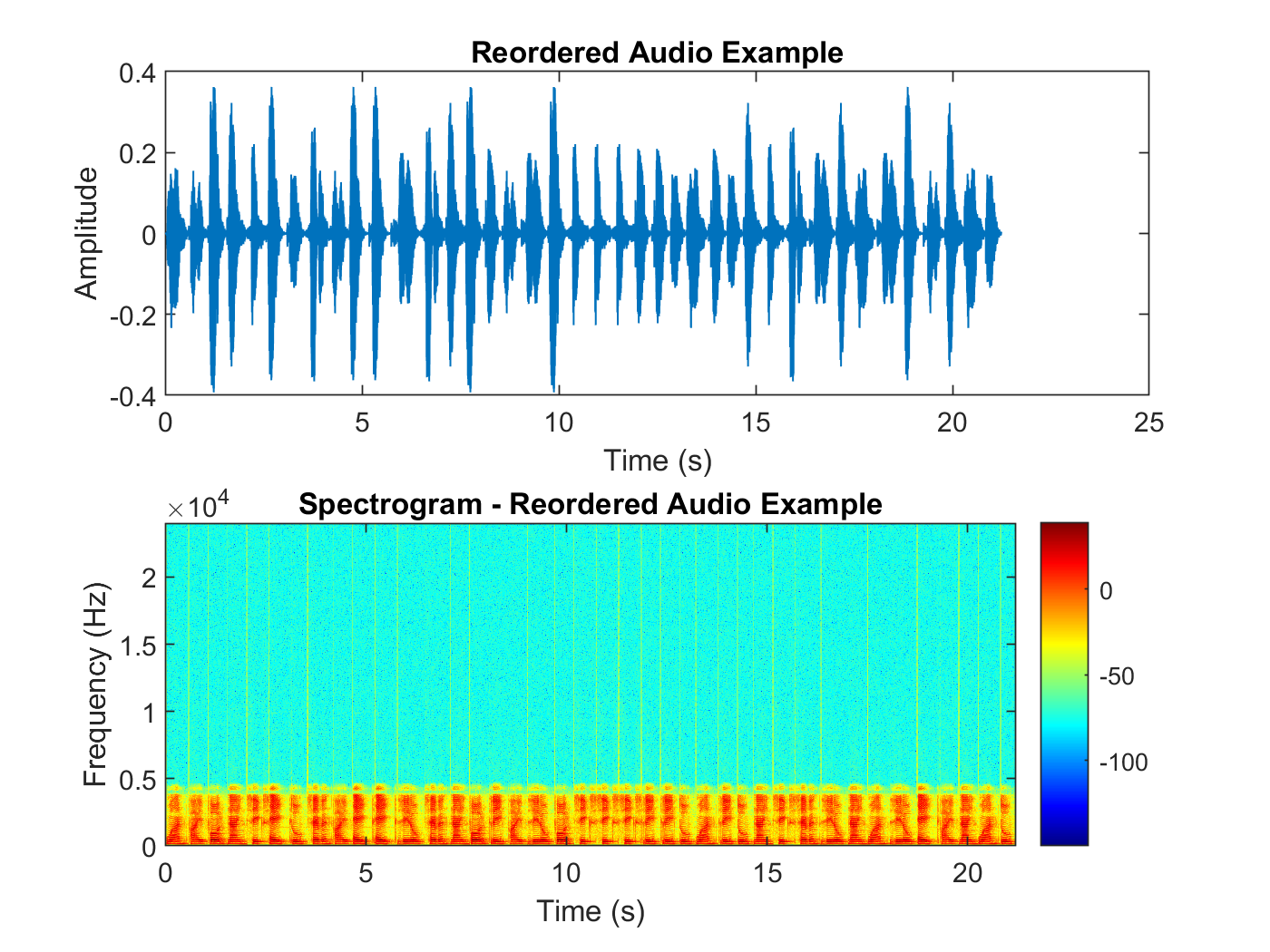}
			\caption{Plot the reordered audio wave in time and frequency domain}
			\label{fig5:plot_time_freq}
			%\vspace{-3mm}
		\end{subfigure}\hspace{35mm}%
		\caption{An example of a speech sample where a host speaks a security code during an authentication ceremony in two different Zoom E2EE sessions.}
		\label{fig4}
		%\vspace{-3mm}
	\end{figure*}

	The attack design is straightforward in the case that the attacker is a legitimate participant in the first meeting, wherein he can listen in on the host's voice and record the numbers from 0 to 9 as they are being spoken. Following the capture of the host's voice, the attacker can subsequently create another Zoom E2EE meeting and carry out a VOICE-ZEUS attack using the previously recorded snippets to generate the new security code in the target host's voice. For instance, if the security code in the first Zoom meeting is (1-2-3-4-5), and the security code in the new Zoom meeting is (5-4-3-2-1), the attacker can utilize the recorded snippets for the first code (1-2-3-4-5) and reorder them to create new audio snippets for the new code (5-4-3-2-1). As described in the Zoom whitepaper \cite{zoomwhitepaper}, the Zoom application utilizes the (\textit{IVK}$_l$) during the handshake protocol, allowing Zoom clients to independently compute the meeting host security code. We found that this security code is generated directly upon the creation and start of a Zoom E2EE meeting, meaning it is available to the attacker before anyone joins the meeting. This gives the attacker a window of opportunity to perform the VOICE-ZEUS attack. To provide an illustration of the results obtained from this experiment, we present Figure \ref{fig4}, which depicts a sample of the security code in the first Zoom meeting and the security code in the new Zoom meeting after reordering the previous host's voice to speak the new code.

	At the same time, we conducted an objective evaluation to quantitatively measure the effectiveness of our attack by assessing the level of similarity between the reordered voice and the original voice. It is typical in speech and speaker recognition systems to acquire a multi-dimensional vector of components from the audio to identify linguistic features. To measure the similarity between a reordered voice that speaks a code and an original voice, we utilized Mel-Cepstral Distortion (MCD), which calculates the Euclidean distance between the feature vectors of the reordered voice and the original voice. Several speech synthesis systems use this method to compare the real and synthetic versions of the same utterance \cite{kominek2008synthesizer,desai2009voice}. If the difference between the feature vector of the original voice and the reordered voice is small enough, the reordered voice will sound quite similar to the original, making it intrinsically difficult to detect the attack. Therefore, a lower MCD indicates a better match between the reordered voice and the original voice, making it challenging to distinguish between the two.

	Computing MCD involves extracting features from the reordered voice, created by remixing previously recorded snippets, and the original voice spoken by the previous host, and subsequently computing the difference between them. The computation of MCD is defined by the following equation:
	
	{
		\small{
			\begin{equation}
				MCD(v^{Reordered}, v^{Original}) = \frac{10}{ln(10)}\sqrt{2\sum_{d=1}^{24}(v_{d}^{Reordered} - v_{d}^{Original})^{2}}
		\end{equation}}
		
	}
	
	\noindent where $v_{d}^{Reordered}$ is the $d$th MCEP\footnote{The weighted average of the magnitudes of cepstral peaks.} of the reordered voice and $v_{d}^{Original}$ is the $d$th MCEP of the original voice. The mel-frequency scaled cepstral coefficients are commonly utilized in various audio and speech processing applications, and they typically consist of parameters ranging from 0 to 24 ($d$ = 0 to 24). The $0$th dimension is typically reserved for capturing the overall signal power or loudness, and as a result, it is crucial to minimize the impact of loudness on the speaker during MCD computation. To accomplish this, MCD can be calculated for dimensions $d$ = 1 to 24, excluding the $0$th dimension used to account for the effect of speaker loudness.
	
	In Tables \ref{tab1} and \ref{tab2}, we present the findings of our objective evaluation. The results reported in Table \ref{tab1} include the outcomes obtained when TTS samples of native English speakers were utilized to record their voices during the authentication ceremony. On the other hand, the results reported in Table \ref{tab2} encompass the outcomes obtained when real human speakers were used to record their voices during the authentication ceremony. Our methodology involved recording the voice of an English speaker during the first Zoom meeting and reordering the spoken numbers in two different audio files. We refer to the first audio file as the (original voice) of the victim speaking the security code in the first meeting and the second audio file as the (reordered voice) generated by remixing the recorded snippets to fit the new security code in the subsequent meeting. We repeated these steps for every speaker in our study. We then calculated the MCD for each speaker. We note that the MCD between the reordered voice and the original voice for each speaker was 0.00, indicating zero dissimilarity between the original and reordered voice in terms of feature vectors. This outcome highlights the effectiveness of the attack, as the reordered voice retains the same features as the original voice despite the applied changes in code digits. This occurs because the attacker reorders previously recorded snippets from voice recordings of the same host to generate the new code, preserving the same voice filtering features as the original voice.

	%First table
	\begin{table*}[t]
		\scriptsize
		\centering
		\caption{Objective Evaluation Results for the VOICE-ZEUS Attack Using TTS Samples}
		\label{tab1}
		%\vspace{-3mm}
		\begin{tabular}{l|cccc|cccc|cccc|}
			\cline{2-13}
			\textbf{}                                                                                                                                       & \multicolumn{2}{c|}{\textbf{Male 1}}                                                                                    & \multicolumn{2}{c|}{\textbf{Female 1}}                                                             & \multicolumn{2}{c|}{\textbf{Male 2}}                                                                                    & \multicolumn{2}{c|}{\textbf{Female 2}}                                                             & \multicolumn{2}{c|}{\textbf{Male 3}}                                                                                    & \multicolumn{2}{c|}{\textbf{Female 3}}                                                             \\ \hline
			\multicolumn{1}{|l|}{\textbf{\begin{tabular}[c]{@{}l@{}}Speaker’s\\ Accent\end{tabular}}}                                                       & \multicolumn{2}{c|}{US English}                                                                                         & \multicolumn{2}{c|}{US English}                                                                    & \multicolumn{2}{c|}{British English}                                                                                    & \multicolumn{2}{c|}{British English}                                                               & \multicolumn{2}{c|}{Australian English}                                                                                 & \multicolumn{2}{c|}{Australian English}                                                            \\ \hline
			\multicolumn{1}{|l|}{\textbf{\begin{tabular}[c]{@{}l@{}}Original\\ Security\\ Code\end{tabular}}}                                               & \multicolumn{4}{c|}{\begin{tabular}[c]{@{}c@{}}40008 55812 36183 49046\\ 68787 34111 34371 14180\end{tabular}}                                                                                                               & \multicolumn{4}{c|}{\begin{tabular}[c]{@{}c@{}}71473 01124 17618 89972\\ 05385 35076 05608 12065\end{tabular}}                                                                                                               & \multicolumn{4}{c|}{\begin{tabular}[c]{@{}c@{}}25088 17100 06194 63649\\ 83034 21827 79601 84031\end{tabular}}                                                                                                               \\ \hline
			\multicolumn{1}{|l|}{\textbf{\begin{tabular}[c]{@{}l@{}}Reordered\\ Security\\ Code\end{tabular}}}                                              & \multicolumn{4}{c|}{\begin{tabular}[c]{@{}c@{}}15496 82758 80794 35046\\ 66332 13296 70910 85912\end{tabular}}                                                                                                               & \multicolumn{4}{c|}{\begin{tabular}[c]{@{}c@{}}44618 84509 13996 12990\\ 53598 89281 53073 85236\end{tabular}}                                                                                                               & \multicolumn{4}{c|}{\begin{tabular}[c]{@{}c@{}}14815 60480 57325 94843\\ 74452 98530 15305 10392\end{tabular}}                                                                                                               \\ \hline
			\multicolumn{1}{|l|}{\textbf{\begin{tabular}[c]{@{}l@{}}Security\\ Code Size\end{tabular}}}                                                     & \multicolumn{2}{c|}{40 Digits}                                                                                          & \multicolumn{2}{c|}{40 Digits}                                                                     & \multicolumn{2}{c|}{40 Digits}                                                                                          & \multicolumn{2}{c|}{40 Digits}                                                                     & \multicolumn{2}{c|}{40 Digits}                                                                                          & \multicolumn{2}{c|}{40 Digits}                                                                     \\ \hline
			\multicolumn{1}{|l|}{\textbf{\begin{tabular}[c]{@{}l@{}}Recording\\ Tool\end{tabular}}}                                                         & \multicolumn{1}{c|}{Audacity} & \multicolumn{1}{c|}{\begin{tabular}[c]{@{}c@{}}Samsung\\ Voice\\ Recorder\end{tabular}} & \multicolumn{1}{c|}{Audacity} & \begin{tabular}[c]{@{}c@{}}Samsung\\ Voice\\ Recorder\end{tabular} & \multicolumn{1}{c|}{Audacity} & \multicolumn{1}{c|}{\begin{tabular}[c]{@{}c@{}}Samsung\\ Voice\\ Recorder\end{tabular}} & \multicolumn{1}{c|}{Audacity} & \begin{tabular}[c]{@{}c@{}}Samsung\\ Voice\\ Recorder\end{tabular} & \multicolumn{1}{c|}{Audacity} & \multicolumn{1}{c|}{\begin{tabular}[c]{@{}c@{}}Samsung\\ Voice\\ Recorder\end{tabular}} & \multicolumn{1}{c|}{Audacity} & \begin{tabular}[c]{@{}c@{}}Samsung\\ Voice\\ Recorder\end{tabular} \\ \hline
			\multicolumn{1}{|l|}{\textbf{\begin{tabular}[c]{@{}l@{}}MCD between\\ the Reordered\\ Voice code and\\ the Original\\ Voice code\end{tabular}}} & \multicolumn{1}{c|}{0.00}     & \multicolumn{1}{c|}{0.00}                                                               & \multicolumn{1}{c|}{0.00}     & 0.00                                                               & \multicolumn{1}{c|}{0.00}     & \multicolumn{1}{c|}{0.00}                                                               & \multicolumn{1}{c|}{0.00}     & 0.00                                                               & \multicolumn{1}{c|}{0.00}     & \multicolumn{1}{c|}{0.00}                                                               & \multicolumn{1}{c|}{0.00}     & 0.00                                                               \\ \hline
		\end{tabular}
	\end{table*}

	%Second table
	\begin{table*}[t]
		\scriptsize
		\centering
		\caption{Objective Evaluation Results for the VOICE-ZEUS Attack Using Human Speech Samples}
		\label{tab2}
		%\vspace{-3mm}
		\begin{tabular}{l|cccc|cccc|cccc|}
			\cline{2-13}
			\textbf{}                                                                                                                                       & \multicolumn{2}{c|}{\textbf{Speaker 1}}                                                                                 & \multicolumn{2}{c|}{\textbf{Speaker 2}}                                                            & \multicolumn{2}{c|}{\textbf{Speaker 3}}                                                                                 & \multicolumn{2}{c|}{\textbf{Speaker 4}}                                                            & \multicolumn{2}{c|}{\textbf{Speaker 5}}                                                                                 & \multicolumn{2}{c|}{\textbf{Speaker 6}}                                                            \\ \hline
			\multicolumn{1}{|l|}{\textbf{\begin{tabular}[c]{@{}l@{}}Speaker’s\\ Language\end{tabular}}}                                                     & \multicolumn{2}{c|}{English}                                                                                            & \multicolumn{2}{c|}{English}                                                                       & \multicolumn{2}{c|}{English}                                                                                            & \multicolumn{2}{c|}{English}                                                                       & \multicolumn{2}{c|}{English}                                                                                            & \multicolumn{2}{c|}{English}                                                                       \\ \hline
			\multicolumn{1}{|l|}{\textbf{\begin{tabular}[c]{@{}l@{}}Original\\ Security\\ Code\end{tabular}}}                                               & \multicolumn{4}{c|}{\begin{tabular}[c]{@{}c@{}}40008 55812 36183 49046\\ 68787 34111 34371 14180\end{tabular}}                                                                                                               & \multicolumn{4}{c|}{\begin{tabular}[c]{@{}c@{}}71473 01124 17618 89972\\ 05385 35076 05608 12065\end{tabular}}                                                                                                               & \multicolumn{4}{c|}{\begin{tabular}[c]{@{}c@{}}25088 17100 06194 63649\\ 83034 21827 79601 84031\end{tabular}}                                                                                                               \\ \hline
			\multicolumn{1}{|l|}{\textbf{\begin{tabular}[c]{@{}l@{}}Reordered\\ Security\\ Code\end{tabular}}}                                              & \multicolumn{4}{c|}{\begin{tabular}[c]{@{}c@{}}15496 82758 80794 35046\\ 66332 13296 70910 85912\end{tabular}}                                                                                                               & \multicolumn{4}{c|}{\begin{tabular}[c]{@{}c@{}}44618 84509 13996 12990\\ 53598 89281 53073 85236\end{tabular}}                                                                                                               & \multicolumn{4}{c|}{\begin{tabular}[c]{@{}c@{}}14815 60480 57325 94843\\ 74452 98530 15305 10392\end{tabular}}                                                                                                               \\ \hline
			\multicolumn{1}{|l|}{\textbf{\begin{tabular}[c]{@{}l@{}}Security\\ Code Size\end{tabular}}}                                                     & \multicolumn{2}{c|}{40 Digits}                                                                                          & \multicolumn{2}{c|}{40 Digits}                                                                     & \multicolumn{2}{c|}{40 Digits}                                                                                          & \multicolumn{2}{c|}{40 Digits}                                                                     & \multicolumn{2}{c|}{40 Digits}                                                                                          & \multicolumn{2}{c|}{40 Digits}                                                                     \\ \hline
			\multicolumn{1}{|l|}{\textbf{\begin{tabular}[c]{@{}l@{}}Recording\\ Tool\end{tabular}}}                                                         & \multicolumn{1}{c|}{Audacity} & \multicolumn{1}{c|}{\begin{tabular}[c]{@{}c@{}}Samsung\\ Voice\\ Recorder\end{tabular}} & \multicolumn{1}{c|}{Audacity} & \begin{tabular}[c]{@{}c@{}}Samsung\\ Voice\\ Recorder\end{tabular} & \multicolumn{1}{c|}{Audacity} & \multicolumn{1}{c|}{\begin{tabular}[c]{@{}c@{}}Samsung\\ Voice\\ Recorder\end{tabular}} & \multicolumn{1}{c|}{Audacity} & \begin{tabular}[c]{@{}c@{}}Samsung\\ Voice\\ Recorder\end{tabular} & \multicolumn{1}{c|}{Audacity} & \multicolumn{1}{c|}{\begin{tabular}[c]{@{}c@{}}Samsung\\ Voice\\ Recorder\end{tabular}} & \multicolumn{1}{c|}{Audacity} & \begin{tabular}[c]{@{}c@{}}Samsung\\ Voice\\ Recorder\end{tabular} \\ \hline
			\multicolumn{1}{|l|}{\textbf{\begin{tabular}[c]{@{}l@{}}MCD between\\ the Reordered\\ Voice code and\\ the Original\\ Voice code\end{tabular}}} & \multicolumn{1}{c|}{0.00}     & \multicolumn{1}{c|}{0.00}                                                               & \multicolumn{1}{c|}{0.00}     & 0.00                                                               & \multicolumn{1}{c|}{0.00}     & \multicolumn{1}{c|}{0.00}                                                               & \multicolumn{1}{c|}{0.00}     & 0.00                                                               & \multicolumn{1}{c|}{0.00}     & \multicolumn{1}{c|}{0.00}                                                               & \multicolumn{1}{c|}{0.00}     & 0.00                                                               \\ \hline
		\end{tabular}
	\end{table*}

	\section{Discussion}
	%\vspace{-3mm}
	\subsection{Summary of Results}
	%\vspace{-1mm}
	Our study demonstrates how easily someone can impersonate a Zoom meeting host by employing a voice channel and conducting a VOICE-ZEUS attack to speak the security code. Despite Zoom expressing concerns about the potential use of deepfake technology for this type of attack \cite{zoomwhitepaper}, we demonstrate that we can compromise the system without employing such sophisticated techniques or attacks. Therefore, our attack surpasses a deepfake approach, delivering superior results at a remarkably low cost and with greater ease of deployment and effectiveness. This flaw in the authentication ceremony significantly compromises the security of Zoom's E2EE mechanism. The attacker could be a malicious participant in a meeting or even a malicious Zoom server. The attack involves collecting audio snippets of all ten digits (from 0 to 9) spoken by the victim. These snippets can be obtained by either a legitimate participant in previous meetings or retrieved from uploaded recorded meetings in the cloud. Collecting ten-digit audio snippets is an effortless task as only a few seconds of eavesdropping can provide sufficient audio samples. During the authentication ceremony, the attacker can wait for the victim to read the security code, which contains all 10 digits. Alternatively, the attacker can record numeric utterances in prior unprotected Zoom meetings. In some cases, the attacker could physically be close to the victim during a public talk or presentation, using the opportunity to record such samples.
	
	We observed that the attacker can easily impersonate the victim's voice because the reordered voice has no distinguishable difference from the original voice, as demonstrated by our objective evaluation where the MCD was equal to 0. This occurs because the attacker reorders the ten spoken digits of the same victim, keeping the voice filtering features, which provides similar characteristics to the original voice. It is obvious that the current implementation of the authentication ceremony in the Zoom application represents a significant vulnerability that threatens the security of Zoom's E2EE mechanism. Once an attacker successfully exploits this vulnerability, he can engage in various nefarious activities, including zoombombing, posting obscene content, making threats against specific individuals, or disseminating morally objectionable content. Our attack poses a significant threat not only to Zoom but also to any E2EE application relying solely on audio channel authentication. While we demonstrated this on Zoom, we assert that E2EE applications like Viber \cite{viber}, utilizing audio channels for number-based authentication, may also be vulnerable.

	\subsection{Mitigation Strategy}
	The susceptibility of numeric security codes in the Zoom application to simple voice manipulations underscores the need to explore alternative mechanisms for addressing and mitigating this vulnerability. One possible approach could involve displaying the security code only after the host and participants have joined the Zoom E2EE meeting session. Currently, the Zoom authentication ceremony only uses the (\textit{IVK}$_l$) during the handshake protocol to generate the security code for the E2EE meeting session. This gives an attacker enough time to execute the VOICE-ZEUS attack by creating the Zoom E2EE meeting session and starting it before all participants join the meeting. In this scenario, the attacker has enough time to reorder the previous snippets in the previous host's voice to speak a new security code. To mitigate this, the Zoom application should generate the security code only after the host and participants have joined the Zoom E2EE meeting session. This would make it challenging for the attacker to have sufficient time to reorder any numbers during the authentication ceremony.
	
	Another potential defense against impersonation attacks is to perform the authentication ceremony using multiple channels, incorporating an additional OOB channel, such as SMS or email, alongside the audio channel. By utilizing multiple channels, the robustness of the system's defense could be enhanced, making it more resistant to manipulation. This would pose a higher level of difficulty for attackers, as they would need to compromise all audio and other OOB channels, effectively doubling their efforts to carry out an impersonation attack. It is important to note that further research and optimization of the implementation of these defense mechanisms are crucial to ensure their effectiveness in mitigating the vulnerability posed by simple voice manipulations in the Zoom application. Careful consideration and thorough testing are necessary to design robust and secure mechanisms that can effectively protect the security of Zoom's E2EE implementation against potential attacks. By adopting these alternative mechanisms and conducting rigorous research and testing, the Zoom application can strengthen its security measures and safeguard against impersonation attacks, thus ensuring the privacy and security of its users.

	\section{Conclusion}
	%\vspace{-1mm}
	Our study of the authentication ceremony and infrastructure of the Zoom application has identified a vulnerability that could compromise the security of E2EE in the presence of malicious actors. Specifically, our study reveals weaknesses in the implementation of group authentication ceremonies in Zoom that make it susceptible to a VOICE-ZEUS attack. We test the likelihood of this attack in two scenarios: one where an attacker is a malicious participant and another where the attacker is a malicious Zoom server with control over the server infrastructure and cloud providers. Our findings demonstrate the feasibility and effectiveness of this attack. Specifically, our simulation experiments reveal that the underlying authentication ceremony of Zoom is susceptible to simple voice manipulations by malicious participants and the malicious Zoom. Exploiting this vulnerability would allow attackers to generate a new security code, thereby jeopardizing the security of future Zoom E2EE meetings and potentially performing zoombombing attacks with offensive content after having impersonated a legitimate host. To mitigate these risks, stronger security measures are necessary during the group authentication ceremony in Zoom. Future research should prioritize the development of effective solutions to mitigate the identified vulnerability, thereby enhancing the overall security of E2EE applications. The findings of our study may contribute to the academic literature on security in video conference systems and provide practical insights for the development of more secure authentication ceremonies. The results could also inform the design and implementation of security measures in other video conference systems beyond Zoom, helping to improve the overall security posture of online communication platforms.
	
	\bibliographystyle{plain}
	\bibliography{references}
	
	%%%%%%%%%%%%%%%%%%%%%%%%%%%%%%%%%%%%%%%%%%%%%%%%%%%%%%%%%%%%%%%%%%%%%%%%%%%%%%%%
\end{document}